\documentclass{sig-alternate-05-2015}
\usepackage{hyperref}
\usepackage{multirow}

\begin{document}

\setcopyright{rightsretained}

\doi{10.475/123_4}

\isbn{123-4567-24-567/08/06}

\acmPrice{\$15.00}

%


\title{Pricing the Woman Card: Gender Politics between Hillary Clinton and Donald Trump}
%
%
%
%
%

\numberofauthors{6} 
%
\author{
%
%
\alignauthor
{Yu Wang}\\
	   \affaddr{Political Science}\\
       \affaddr{University of Rochester}\\
       \affaddr{Rochester, NY, 14627}\\
       \email{yu.wang@rochester.edu}
         \alignauthor
      Yang Feng\\
       \affaddr{Computer Science}\\
       \affaddr{University of Rochester}\\
       \affaddr{Rochester, NY, 14627}\\
       \email{yfeng23@cs.rochester.edu}
            \alignauthor
Yuncheng Li\\
	   \affaddr{Computer Science}\\
       \affaddr{University of Rochester}\\
       \affaddr{Rochester, NY, 14627}\\
       \email{yli@cs.rochester.edu}   \\
       \and  
       \alignauthor        
       Xiyang Zhang\\
	   \affaddr{Psychology}\\
       \affaddr{University of Oklahoma}\\
       \affaddr{Norman, OK, 73019}\\
       \email{xiyang.zhang-1@ou.edu}
       \alignauthor   
       Richard Niemi\\
	   \affaddr{Political Science}\\
       \affaddr{University of Rochester}\\
       \affaddr{Rochester, NY, 14627}\\
       \email{niemi@rochester.edu}      
         \alignauthor
         Jiebo Luo\\
       \affaddr{Computer Science}\\
       \affaddr{University of Rochester}\\
       \affaddr{Rochester, NY, 14627}\\
       \email{jluo@cs.rochester.edu}  
}

\maketitle
\begin{abstract}
In this paper, we propose a data-driven method to measure the impact of the `woman card' exchange between Hillary Clinton and Donald Trump. Building from a unique dataset of the two candidates' Twitter followers, we first examine the transition dynamics of the two candidates' Twitter followers one week before the exchange and one week after. Then we train a convolutional neural network to classify the gender of the followers and unfollowers, and study how women in particular are reacting to the `woman card' exchange. Our study suggests that the `woman card' comment has made women more likely to follow Hillary Clinton, less likely to unfollow her and that it has apparently not affected the gender composition of Trump followers.
\end{abstract}

%
%
%

\ccsdesc[500]{Human-centered computing~Social engineering (social sciences)}
\ccsdesc[500]{Human-centered computing~Social media}
\printccsdesc

\keywords{Presidential Election; Donald Trump; Hillary Clinton; Gender; Woman Card;}

\section{Introduction}
During his victory speech on April 26, 2016, Donald Trump accused Hillary Clinton of playing the `woman card,' and said that she would be a failed candidate if she were a man. Clinton fired back during her victory speech in Philadelphia and said that ``If fighting for women's health care and paid family leave and equal pay is playing the `woman card,' then deal me in.'' The `woman card' subsequently became the meme of the week and its effects are much debated. According to \textit{CNN, New York Times, Washington Post and The Financial Times}, this exchange between the two presidential front-runners signaled a heated general election clash over gender.\footnote{See for example, \url{http://www.nytimes.com/2016/04/29/us/politics/hillary-clinton-donald-trump-women.html.}} 

\begin{figure}[!h]
\centering
\includegraphics[height=5.5cm,width=8.4cm]{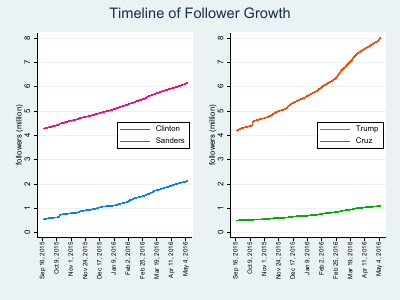}
\caption{Donald Trump and Hillary Clinton lead in terms of Twitter followers.}
\label{timeline}
\end{figure}

In this paper, we propose a data-driven method to measure the impact of the `woman card' exchange between Hillary Clinton and Donald Trump. Building from a unique dataset of the two candidates' Twitter followers, we first examine the transition dynamics of the two candidates' Twitter followers one week before the woman's card controversy and one week after. Then we train a convolutional neural network to classify the followers' gender and study how women in particular are reacting to the `woman card' exchange. Our study suggests that the `woman card' comment has made women more likely to follow Hillary Clinton, less likely to unfollow her and that apparently it has not affected the gender composition of Trump followers.

\section{Related Literature}
Our work builds on previous literature in electoral studies, data mining, and computer vision.

In eletoral studies, researchers have argued that gender constitutes an important factor in voting behavior. One common observation is that women tend to vote for women, which is usually referred to as gender affinity effect \cite{sexAndGOP,genderAffinityEffect}. In the 2016 presidential election, Hillary Clinton also portrays herself as a champion ``fighting for women's healthcare and paid family leave and equal pay.'' Our work will test the strength of this gender affinity effect.

In data mining, there is a burgeoning literature on using social media data to analyze and predict elections. In particular, several studies have explored ways to infer users' preferences. According to \cite{tweets2polls}, tweets with sentiment can potentially serve as votes and substitute traditional polling. \cite{trumponfire} exploits the variations in the number of `likes' of the tweets to infer Trump followers' topic preferences. \cite{facebookCongress} uses candidates' ``likes" in Facebook to quantify a campaign's success in engaging the public. \cite{neco} uses follower growth on public debate dates to measure candidates' debate performance. Our work also pays close attention to the number of followers, but we go further by examining both new followers and unfollowers.

Our work also ties in with current computer vision research. In this dimension, our work is related to gender classification using facial features. \cite{israel} uses a five-layer network to classify both age and gender. \cite{ginosar} introduces a dataset of frontal-facing American high school yearbook photos and uses the extracted facial features to study historical trends. Their work is the inspiration of ours. \cite{facerace} provides a comprehensive survey of race classification based on facial features. \cite{trumpists} uses user profile images to study and compare the social demographics of Trump followers and Clinton followers. \cite{votingfeet} focuses specifically on the unfollowers of the Donald Trump and Hillary Clinton. Here our work goes one step further and looks at the new followers. In addition, our work focuses exclusively on the `woman card.'

\section{Data and Methodology}
In this section, we describe our dataset \textit{US2016}, the pre-processing procedures and our CNN model. One key variable is \textit{number of followers.} This variable is available for both candidates and covers the entire period from Sept. 18, 2015 to May 7, 2016. The two presidential front-runners also have by far the most Twitter followers (Figure \ref{timeline}). This variable is updated every 10 minutes. 

Besides the number of followers, our dataset \textit{US2016} also  contains the detailed IDs of Trump's and Clinton's followers. Specifically for this paper, we are able to use these IDs to identify all the new followers and the unfollowers of Donald Trump first between April 19 and April 26 and then between April 26 and May 1. Similarly, we have information on Hillary Clinton's new followers and unfollowers first between April 20 and April 27 and then between April 27 and May 2. This enables us to examine in a definitive manner the gender composition of new followers and unfollowers one week before the `woman card' exchange (April 26) and one week after. We report the summary statistics in Table \ref{sum-id}.

\begin{table}[!h]
\centering
\setlength{\tabcolsep}{7.5pt}
\caption{Mobility in the Candidates' Followers}
\label{sum-id}
\begin{tabular}{@{\extracolsep{4pt}}lllll@{}}
\hline\hline
        \multirow{2}{*}{`Woman Card'  }      & \multicolumn{2}{c}{Hillary Clinton} & \multicolumn{2}{c}{Donald Trump} \\\cline{2-3}\cline{4-5}
          & Before            & After           & Before          & After          \\\hline
New Followers & 72266             & 54820           & 116456          & 115246         \\\hline
Unfollowers   & 9572              & 8393            & 18376           & 18292 \\ 
\hline       
\end{tabular}
\end{table}
Furthermore, as we have the follower information of other presidential candidates such as Bernie Sanders and Ted Cruz, we are able to identify the destinations of Trump and Clinton unfollowers. We report these statistics in Table \ref{Clinton-unfollow} and Table \ref{Trump-unfollow}.

\begin{table}[!h]
\centering
\setlength{\tabcolsep}{4pt}
\caption{Mobility of Hillary Clinton's Unfollowers}
\label{Clinton-unfollow}
\begin{tabular}{lccl}
\hline\hline
Destination & Bernie Sanders & Donald Trump & Ted Cruz* \\\hline
 Before     & 14.55\%        & 11.95\%      & 2.19\% \\ \hline
  After     & 12.47\%        & 11.03\%      & 2.62\% \\ \hline
\multicolumn{4}{l}{*Ted Cruz has dropped out after the Indiana primary.}\\
\end{tabular}
\end{table}

\begin{table}[!h]
\centering
\setlength{\tabcolsep}{4pt}
\caption{Mobility of Donald Trump's Unfollowers}
\label{Trump-unfollow}
\begin{tabular}{lccl}
\hline\hline
Destination & Hillary Clinton & Bernie Sanders & Ted Cruz \\\hline
 Before     & 6.04\%        & 4.87\%      & 4.55\% \\ \hline
  After     & 5.94\%        & 4.54\%      & 3.70\% \\ \hline
\end{tabular}
\end{table}

We collect data on the profile images based on follower (unfollower) IDs, and our goal is to infer an individual's gender based on the profile image.

To process the profile images, we first use OpenCV to identify faces, as the majority of profile images only contain a face.\footnote{http://opencv.org/.} We discard images that do not contain a face and the ones in which OpenCV is not able to detect a face. When multiple faces are available, we choose the largest one. Out of all facial images thus obtained, we select only the large ones. Here we set the threshold to 18kb. This ensures high image quality and also helps remove empty faces. Lastly we resize those images to (28, 28, 3). In Table \ref{image}, we report the summary statistics of the images used in classification.

\begin{table}[!h]
\centering
\caption{Number of Profile Images in \textit{US2016}}
\setlength{\tabcolsep}{7.5pt}
\label{image}
\begin{tabular}{@{\extracolsep{4pt}}lllll@{}}
\hline\hline
        \multirow{2}{*}{`Woman Card'  }      & \multicolumn{2}{c}{Hillary Clinton} & \multicolumn{2}{c}{Donald Trump} \\\cline{2-3}\cline{4-5}
            & Before            & After           & Before          & After          \\\hline
New Followers & 14504             & 11147           & 20204          & 21187         \\\hline
Unfollowers   & 2039              & 1587            & 3682            & 3036 \\ 
\hline       
\end{tabular}
\end{table}

To classify profile images, we train a convolutional neural network using 42,554 weakly labeled images, with a gender ratio of 1:1. These images come from Trump's and Clinton's current followers. And we infer their labels using the followers' names. For example, James, John, Luke and Michael are male names, and Caroline, Elizabeth, Emily, Isabella and Maria are female names.\footnote{The list of label names together with the validation data set and the trained model, is now available at the first author's website.} For validation, we use a manually labeled data set of 1,965 profile images for gender classification. The validation images come from Twitter as well, so we can avoid the cross-domain problem. Moreover, they do not intersect with the training samples as they come exclusively from individuals who unfollowed Hillary Clinton before March 2016.

\begin{figure*}[!htb]
\centering
\includegraphics[height=5.5cm,width=15cm]{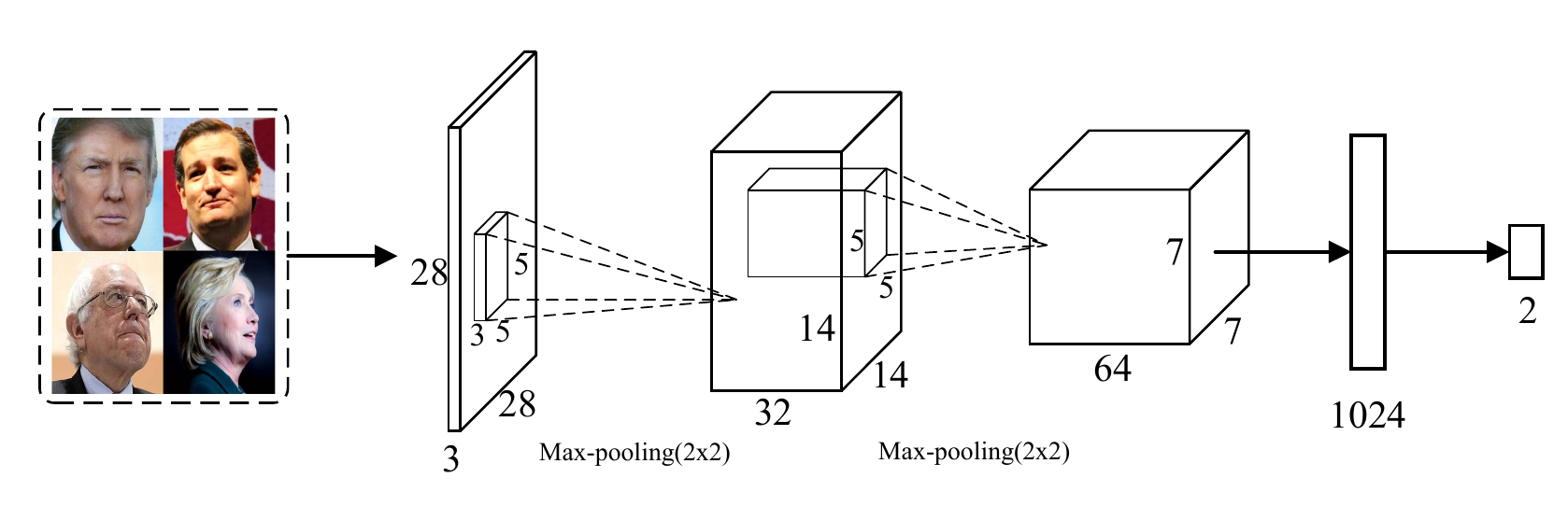}
\caption{The CNN model consists of 2 convolution layers, 2 max-pool layers and a fully connected layer.}
\label{visio}
\end{figure*} 

\begin{table}[!h]
\centering
\caption{Summary Statistics of CNN Performance}
\label{Performance}
\setlength{\tabcolsep}{6.5pt}
\begin{tabular}{lllll}
\hline\hline
Architecture & Precision & Recall & F1    & Accuracy \\
2CONV-1FC    & 91.36     & 90.05  & 90.70 & 90.18   \\
\hline
\end{tabular}
\end{table}

The architecture of our convolutional neural network is illustrated in Figure \ref{visio}, and the performance of the model is reported in Table \ref{Performance}.

\section{Main Results}
In this section, we analyze the effects of the `woman card' exchange on the gender composition of new followers and unfollowers for both Hillary Clinton and Donald Trump. Specifically, we will examine whether this exchange has made women more likely to follow Hillary Clinton and more likely to leave Trump. As reported in Section 3, we have set the time window of observation to one week.

\subsection{New Followers}
\begin{figure}[!h]
\centering
\includegraphics[height=4.5cm,width=8.4cm]{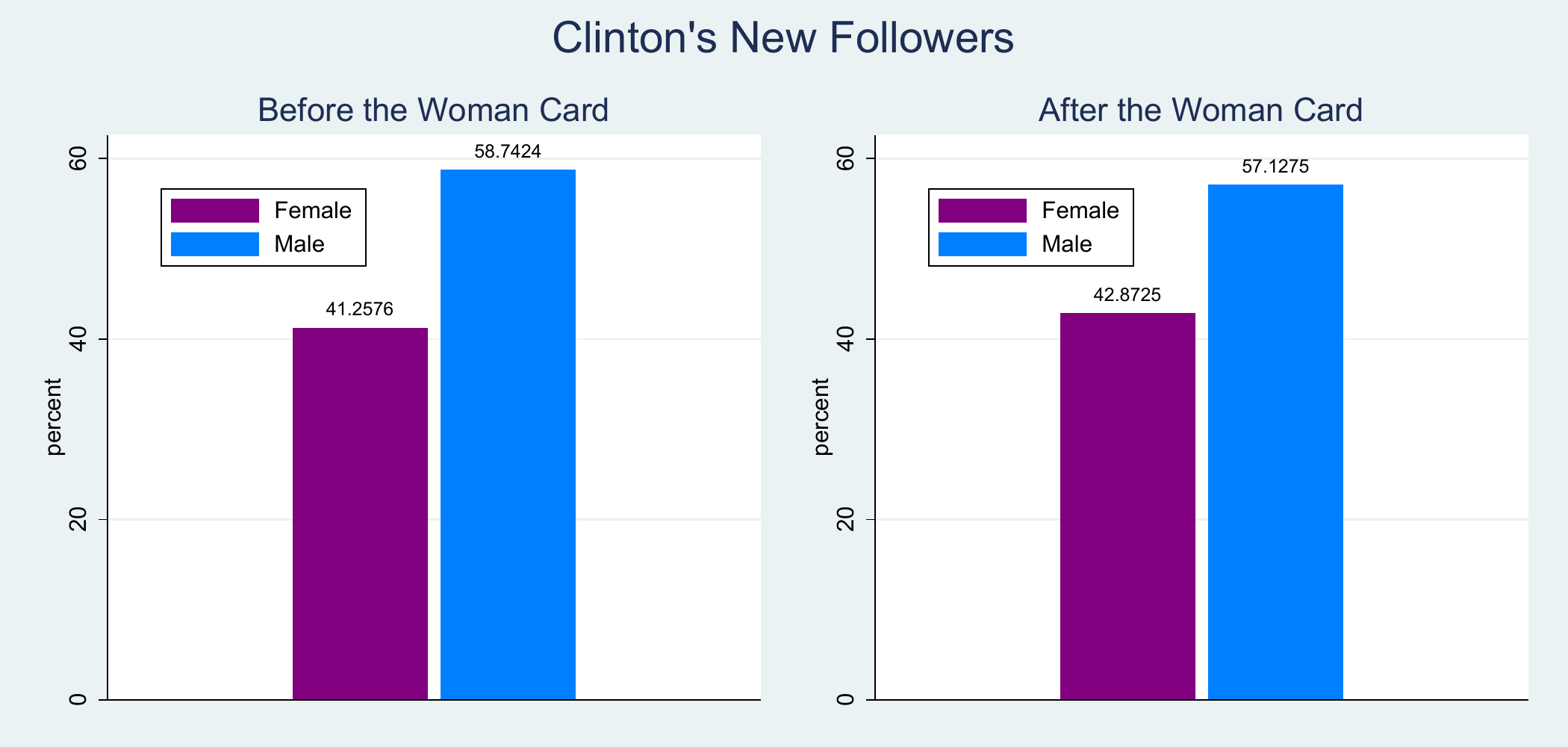}
\caption{Gender Composition of Hillary Clinton's New Followers.}
\label{woman-follow-clinton}
\end{figure}

In Figure \ref{woman-follow-clinton}, we report on the gender composition of Clinton's new followers one week before the `woman card' exchange and one week after. We observe a 1.6\% increase in percentage of women followers. Our sample size is 14,504 for the first week and 11,147 for the second.

\begin{figure}[!h]
\centering
\includegraphics[height=4.5cm,width=8.4cm]{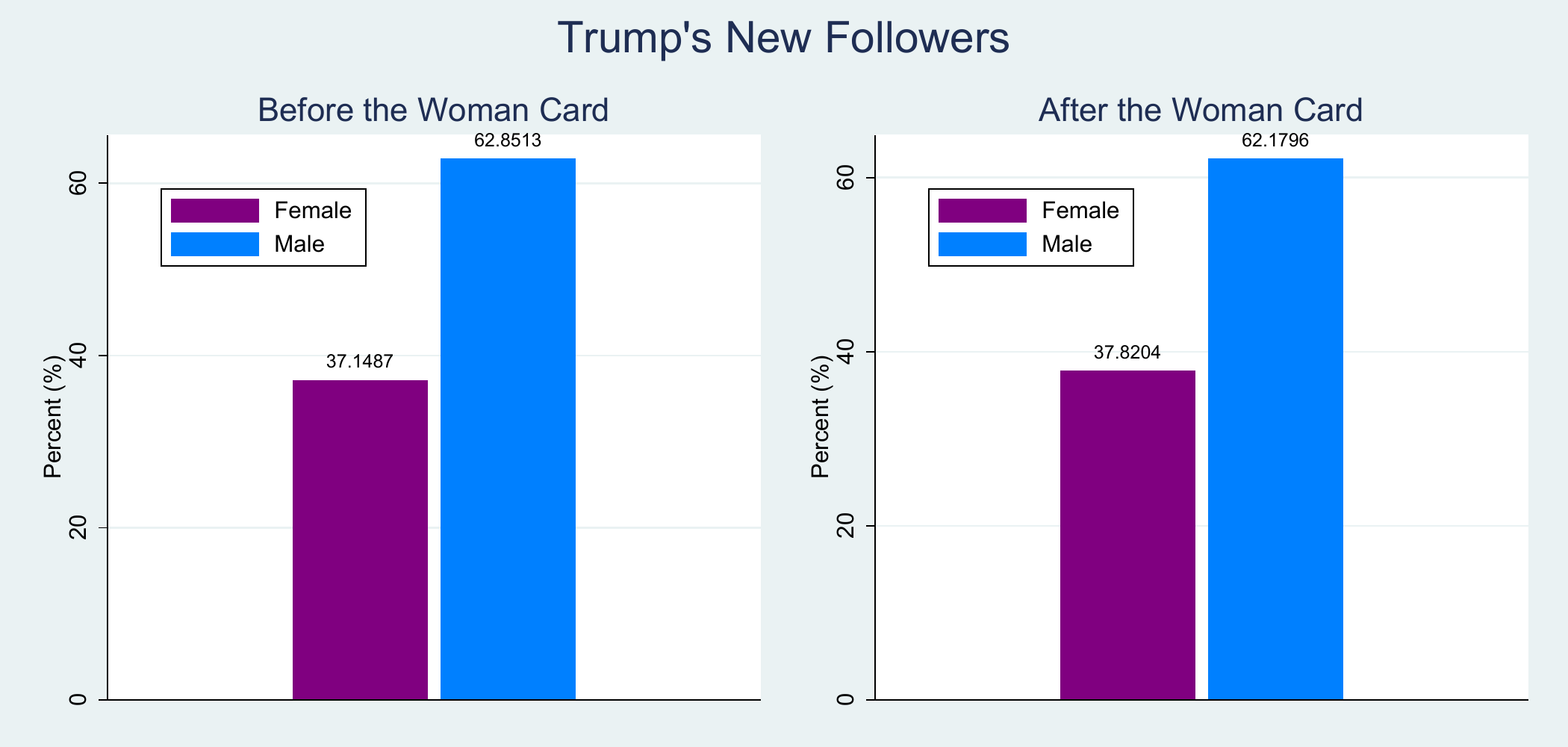}
\caption{Gender Composition of Donald Trump's New Followers.}
\label{woman-follow-trump}
\end{figure}

In Figure \ref{woman-follow-trump}, we report on the gender composition of Trump's new followers one week before the `woman card' exchange and one week after. We observe a 0.6717\% increase in percentage of women followers. Our sample size is 20,204 for the first week and for the second 21,187. While our main focus is the time-series variations for the candidates, it is interesting to note that cross candidates, Clinton attracts more new female followers proportionally than Trump.

Using score test (Table \ref{stgender1}), we are able to show that for Clinton the surge of female presence among her new followers is statistically significant.\footnote{The formula for the score test statistic is: $z=\frac{\hat{p}_1-\hat{p}_2}{\sqrt{\hat{p}(1-\hat{p})(1/n_1+1/n_2)}}$, where $\hat{p}_1=\frac{x}{n_1},\:\hat{p}_2=\frac{y}{n_2},\:p=\frac{x+y}{n_1+n2}.$ With large $n_1$ and $n_2$, z is approximately standard normal.} The same does not hold for Donald Trump.

\begin{table}[h!]
\centering
\caption{New Followers' Gender Composition}\label{stgender1}
\setlength{\tabcolsep}{3pt}
\begin{tabular}{lllll}\hline\hline
\multirow{2}{*}{Null Hypothesis}& \multicolumn{2}{c}{Clinton} & \multicolumn{2}{c}{Trump} \\
\cline{2-3}\cline{4-5}
                                 & z statistic     & \textit{p}  value       & z statistic     & \textit{p}  value \\\hline
p$_{before}$=p$_{after}$      & 2.597           &  0.0093                         & 1.411 & 0.1582   \\\hline
\end{tabular}
\end{table}

\subsection{Unfollowers}

\begin{figure}[!h]
\centering
\includegraphics[height=4.5cm,width=8.4cm]{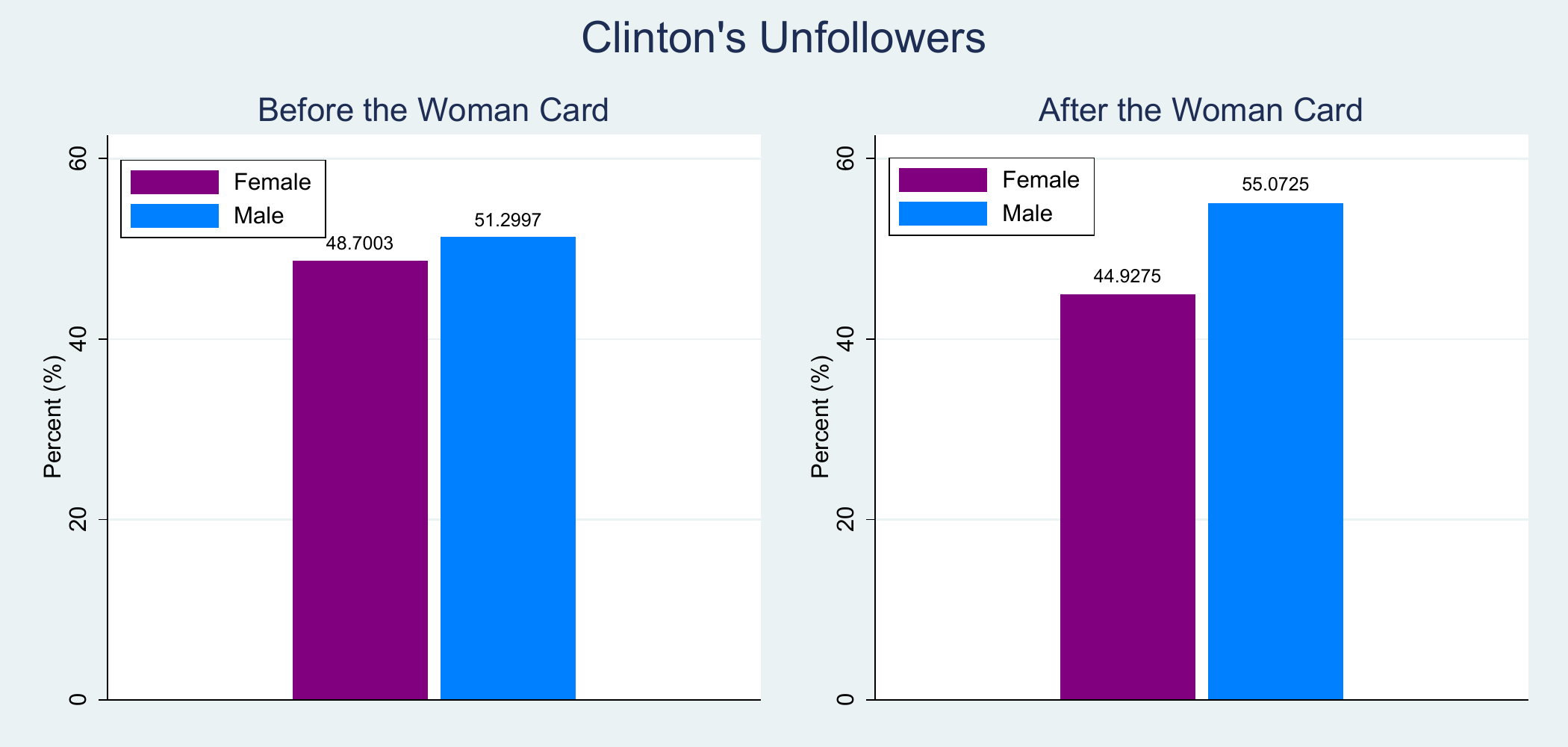}
\caption{Gender Composition of Clinton's Unfollowers.}
\label{woman-unfollow-clinton}
\end{figure}

In Figure \ref{woman-unfollow-clinton}, we report on the gender composition of Clinton's unfollowers one week before the `woman card' exchange and one week after. We observe a 3.7728\% decrease in the percentage of women unfollowers. Our sample size is 2,039 for the first week and 1,587 for the second.

\begin{figure}[!h]
\centering
\includegraphics[height=4.5cm,width=8.4cm]{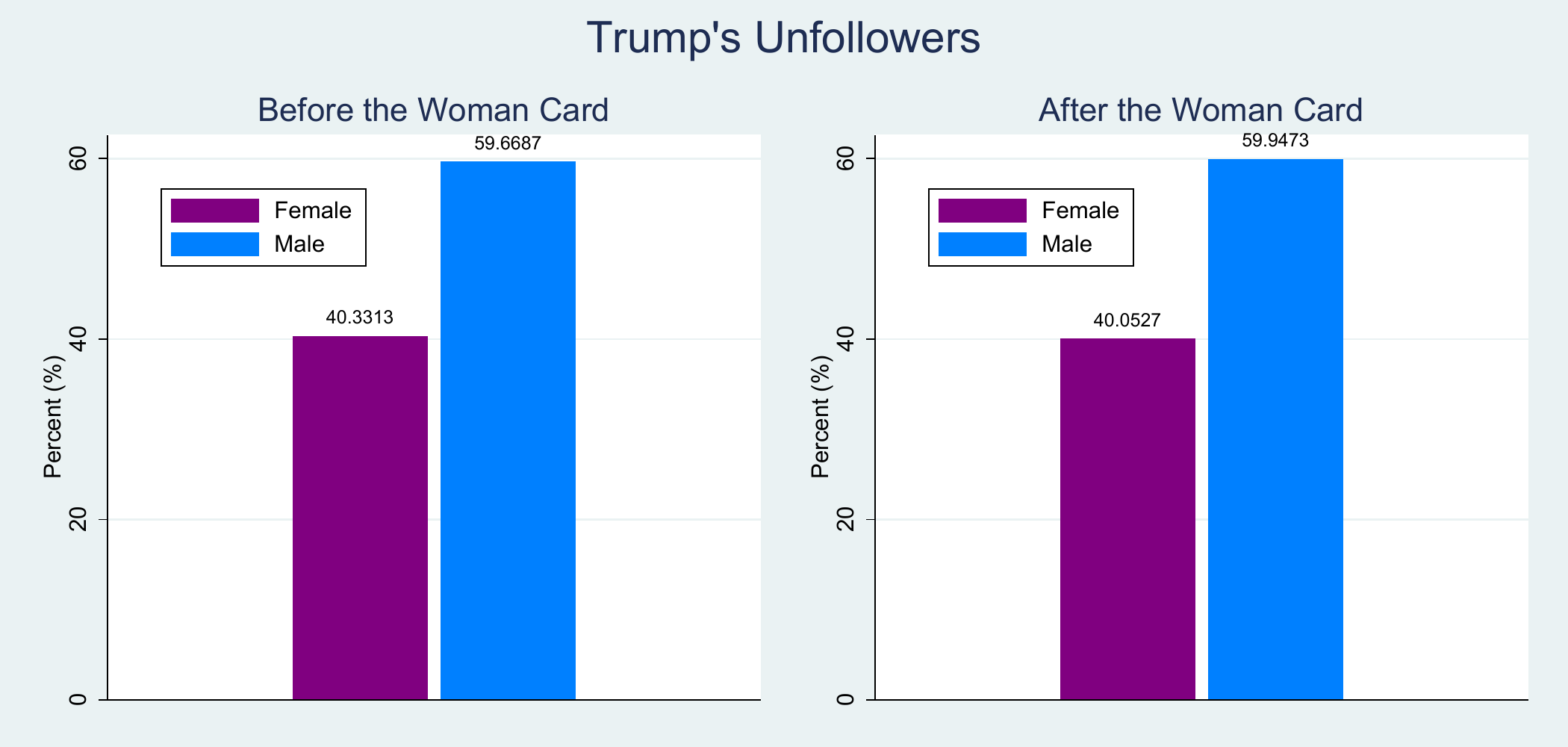}
\caption{Donald Trump and Hillary Clinton lead in terms of Twitter followers.}
\label{woman-unfollow-trump}
\end{figure}

In Figure \ref{woman-unfollow-trump}, we report on the gender composition of Trump's unfollowers one week before the `woman card' exchange and one week after. We observe a 0.2786\% decrease in the percentage of women unfollowers. Our sample size is 3,682 for the first week and 3,036 for the second.

Using score test (Table \ref{stgender}), we show that for Clinton the decrease of female presence among her unfollowers is statistically significant at 95\% confidence interval. While Donald Trump also observes a decrease in the percentage of female unfollowers, the decrease is not statistically significant.

\begin{table}[h!]
\centering
\caption{Unfollowers' Gender Composition}\label{stgender}
\setlength{\tabcolsep}{3pt}
\begin{tabular}{lllll}\hline\hline
\multirow{2}{*}{Null Hypothesis}& \multicolumn{2}{c}{Clinton} & \multicolumn{2}{c}{Trump} \\
\cline{2-3}\cline{4-5}
                                 & z statistic     & \textit{p}  value       & z statistic     & \textit{p}  value \\\hline
p$_{before}$=p$_{after}$      & -2.2581           &  0.0239 & -0.23178 & 0.8167   \\\hline
\end{tabular}
\end{table}

\section{Conclusions}

Gender will play a crucial role in the 2016 U.S. presidential election campaign. In this paper we have proposed a data-driven method to measure the effects of the first episode of the gender war between Hillary Clinton and Donald Trump: the `woman card'. Building from a unique dataset of the two candidates' Twitter followers, we trained a convolutional neural network to classify the gender of followers and unfollowers, and study in particular how women are reacting to the `woman card' exchange.

Our study suggests that the `woman card' exchange has made women both more likely to follow Hillary Clinton and less likely to unfollow her. Equally important, this exchange has apparently not affected the gender composition of Trump followers. Our study has provided the first evidence of the possible impacts of the gender wars between Hillary Clinton and Donald Trump. 

\section{Acknowledgment}
We gratefully acknowledge support from the University and from our corporate sponsors.
\bibliographystyle{acm}
\bibliography{yu}  

\begin{thebibliography}{10}

\bibitem{genderAffinityEffect}
{\sc Dolan, K.}
\newblock {I}s {T}here a ``{G}ender {A}ffinity {E}ffect" in {A}merican
  {P}olitics? {I}nformation, {A}ffect, and {C}andidate {S}ex in {U}.{S}.
  {H}ouse {E}lections.
\newblock {\em Political Research Quarterly\/} (2008).

\bibitem{facerace}
{\sc Fu, S., He, H., and Hou, Z.-G.}
\newblock {L}earning {R}ace from {F}ace: {A} {S}urvey.
\newblock In {\em Pattern Analysis and Machine Intelligence, IEEE Transactions
  on\/} (2014), vol.~36.

\bibitem{ginosar}
{\sc Ginosar, S., Rakelly, K., Sachs, S., Yin, B., and Efros, A.~A.}
\newblock {A} {C}entury of {P}ortraits: {A} {V}isual {H}istorical {R}ecord of
  {A}merican {H}igh {S}chool {Y}earbooks.
\newblock In {\em ICCV 2015 Extreme Imaging Workshop Proceedings\/} (2015).

\bibitem{sexAndGOP}
{\sc King, D.~C., and Matland, R.~E.}
\newblock {S}ex and the {G}rand {O}ld {P}arty: {A}n {E}xperimental
  {I}nvestigation of the {E}ffect of {C}andidate {S}ex on {S}upport for a
  {R}epublican {C}andidate.
\newblock {\em American Politics Research\/} (2003).

\bibitem{israel}
{\sc Levi, G., and Hassner, T.}
\newblock {A}ge and {G}ender {C}lassification using {D}eep {C}onvolutional
  {N}eural {N}etworks.
\newblock In {\em {P}roceedings of the {IEEE} {C}onference on {C}omputer
  {V}ision and {P}attern {R}ecognition\/} (2015), pp.~34--42.

\bibitem{facebookCongress}
{\sc MacWilliams, M.~C.}
\newblock {F}orecasting {C}ongressional {E}lections {U}sing {F}acebook {D}ata.
\newblock {\em PS: Political Science \& Politics 48}, 04 (October 2015).

\bibitem{tweets2polls}
{\sc O'Connor, B., Balasubramanyan, R., Routledge, B.~R., and Smith, N.~A.}
\newblock {F}rom {T}weets to {P}olls: {L}inking {T}ext {S}entiment to {P}ublic
  {O}pinion {T}ime {S}eries.
\newblock In {\em Proceedings of the Fourth International AAAI Conference on
  Weblogs and Social Media\/} (2010).

\bibitem{trumpists}
{\sc Wang, Y., Li, Y., and Luo, J.}
\newblock {D}eciphering the 2016 {U}.{S}. {P}residential {C}ampaign in the
  {T}witter {S}phere: {A} {C}omparison of the {T}rumpists and {C}lintonists.
\newblock In {\em {T}enth {I}nternational {AAAI} {C}onference on {W}eb and
  {S}ocial {M}edia\/} (2016).

\bibitem{votingfeet}
{\sc Wang, Y., Li, Y., You, Q., Zhang, X., Niemi, R., and Luo, J.}
\newblock {V}oting with {F}eet: {W}ho are {L}eaving {H}illary {C}linton and
  {D}onald {T}rump?
\newblock In {\em ar{X}iv preprint :1604.07103\/} (2016).

\bibitem{neco}
{\sc Wang, Y., Luo, J., Niemi, R., and Li, Y.}
\newblock {T}o {F}ollow or {N}ot to {F}ollow: {A}nalyzing the {G}rowth
  {P}atterns of the {T}rumpists on {T}witter.
\newblock In {\em {W}orkshop {P}roceedings of the 10th {I}nternational {AAAI}
  {C}onference on {W}eb and {S}ocial {M}edia\/} (2016).

\bibitem{trumponfire}
{\sc Wang, Y., Luo, J., Niemi, R., Li, Y., and Hu, T.}
\newblock {C}atching {F}ire via `{L}ikes': {I}nferring {T}opic {P}references of
  {T}rump {F}ollowers on {T}witter.
\newblock In {\em {T}enth {I}nternational {AAAI} {C}onference on {W}eb and
  {S}ocial {M}edia\/} (2016).

\end{thebibliography}

\end{document}